\documentclass{ws-ijmpa}

\newcommand{\nn}{\nonumber}
\newcommand{\be}{\begin{equation}}
\newcommand{\ee}{\end{equation}}
\newcommand{\ba}{\begin{eqnarray}}
\newcommand{\ea}{\end{eqnarray}}
\newcommand{\req}[1]{(\ref{#1})}
\def\={\,=\,}
\newcommand{\ci}[1]{\cite{#1}}

\def\mev{~{\rm MeV}}
\def\gev{~{\rm GeV}}

\def\als{\alpha_{\rm s}}
\def\et1{\widetilde{\eta}_1}
\def\et8{\widetilde{\eta}_8}
\def\eps{\epsilon}

\def\etp{\eta^\prime}

\def\ab{\bar{a}}
\def\ub{\bar{u}}
\def\db{\bar{d}}

\begin{document} 
\markboth{Kroll}{Mixing}
%
\catchline{}{}{}{}{}
%
\title{\bf MIXING OF PSEUDOSCALAR MESONS\\ AND \\ISOSPIN SYMMETRY BREAKING}

\author{P.\ KROLL} 

\address{Fachbereich Physik, Universit\"at Wuppertal,\\ 
D-42097 Wuppertal, Germany}

\maketitle

\begin{abstract}
Mixing of the pseudoscalar mesons is discussed in the quark-flavor
basis with the hypothesis that the basis decay constants follow 
the pattern of particle state mixing. The divergences of the axial
vector currents which embody the axial vector anomaly, combined with
this hypothesis provide a calculational scheme for
the parameters describing the mixing of the $\pi^0$, $\eta$
and $\etp$ mesons. Phenomenological applications of this mixing scheme
are presented with particular interest focussed on isospin symmetry
breaking in QCD estimated as $\eta$ and $\eta'$ admixtures to
the pion. In contrast to previous work a possible difference in the
basis decay constants $f_u$ and $f_d$ is considered and consequences
of this potentially large effect on the strength of isospin symmetry
breaking is discussed.
\end{abstract}
\section{Introduction}
The mixing of the pseudoscalar mesons is a subject of considerable
interest that has been examined in many theoretical and experimental
investigations, for earlier references see \ci{isgur}. New aspects of
mixing which mainly concern the proper definition of meson decay
constants and the consistent extraction of mixing parameters from
experimental data have recently been discussed \ci{kaiser,FKS1,FKS2}
and will be reviewed here. An important topic in this context is the
interplay between the U${}_{\rm A}$(1) anomaly and flavor symmetry
breaking.

In the analysis of $\eta-\etp$ mixing presented in Ref.~\ci{FKS1} 
the quark flavor basis~\cite{isgur} is used and assumed that the
decay constants ($f_q$, $q=u,d,s$) in that basis follow the pattern of
particle state mixing, i.e.\ there is a common mixing angle, $\phi$,
in contrast to the frequently used octet-singlet basis where different 
mixing angles for the decay constants and for the states are needed. 
With the help of the divergences of the axial vector currents the
basic parameters of that mixing scheme can be 
determined for given masses of the physical mesons. It has been found
in \cite{FKS1,FKS2} that this approach leads to consistent results and 
explains many empirical features of $\eta-\etp$ mixing. In Ref.\
\ci{FKS2} that mixing scheme has been generalized in order to
estimate the $\eta$ and $\etp$ admixtures to the pion which is a
source of isospin symmetry breaking (ISB). Due to a number of recent experiments 
\ci{BES,tippens:01,opper:03,steph:03,abdel:03} the interest in ISB has been 
renewed. It therefore seems opportune to reinvestigate
$\pi^0-\eta-\etp$ mixing. Thus, for instance, a crucial examination of the
value of the $u-d$ quark mass difference is necessary. This mass
difference is an important ingredient in understanding and estimating
ISB within QCD. Another point is the role of possible differences between the
decay constants $f_u$ and $f_d$ which has not been 
explored so far. Such a contribution to ISB would by no means ruin our
understanding of it as being generated by the $u - d$
quark-mass differences. This is so because decay constants which  
represent wavefunctions at zero spatial quark-antiquark separation,
are functions of the quark masses. Since we are not able to calculate
the decay constants to a sufficient accuracy within QCD at present
one may consider them as independent soft parameters in the 
analysis of meson mixing and ISB. As I am going to explain below a
possible difference between these decay constants is an important
source of theoretical uncertainty in the analysis of ISB.

The plan of the paper is as follows: In Sect.\ 2 the quark-flavor
mixing scheme will be presented briefly. In Sect.\ 3 some theroretical
and phenomenological results for $\eta$-$\etp$ mixing will be
reviewed. Mixing with the $\pi^0$ and consequences for ISB will be
discussed in Sect.\ 4 and compared to experiment  in Sect.\ 5.
The summary is given in Sect.\ 6.
\section{The quark-flavor mixing scheme} 
The quark-flavor basis is constructed by the states $\eta_{\,a}$, $a=u,d,s$ 
which are understood to possess the parton compositions
\be
|\eta_{\,a}\rangle \= \Psi_a \, |a\ab \rangle + \cdots\,
\ee
in a Fock expansion where $\Psi_a$ denotes a (light-cone) wave function. 
The higher Fock states also include $|gg\rangle$ components. The decay
constants in that basis, defined as vacuum-particle matrix elements of 
axial-vector currents, $J^a_{5\mu}=\ab \gamma_\mu \gamma_5\, a$, are
{\it assumed} to possess the property
\be
\langle0\mid J^a_{5\mu}|\eta_{\,b}(p)\rangle \= \delta_{ab}\, f_a\, p_\mu\,,
\qquad a,b=u,d,s\,.
\label{currents}
\ee
The motivation for choosing this specific basis comes from the fact
that breaking of SU(3)${}_{\rm F}$ by the quark masses influences the
three states differently, and from the observation that vector and
tensor mesons - where the axial vector anomaly plays no role - have
state mixing angles very close to the ideal mixing one $\theta_{\rm
  ideal}=\arctan{\sqrt{2}}$. The quark-flavor mixing scheme holds
to the extent that OZI rule violation except of those
induced by the anomaly, are negligible small. Flavor symmetry breaking,
on the other hand, is rather large and is to be taken into account in
any analysis of pseudoscalar meson mixing.

Since mixing of the $\pi^0$ with the $\eta$ and $\etp$ is weak while
$\eta-\etp$ mixing is strong it is appropriate to use isoscalar and 
isovector combinations \footnote{
In Ref.\ \ci{FKS1} $\eta_{\,+}$ is denoted by $\eta_{\,q}$. Analogous
changes occur for other quantities, e.g.\ $f_+ \equiv f_q$.}
\be
\eta_{\,\pm} \= \frac1{\sqrt{2}}\, [\eta_{\,u}\pm \eta_{\,d}] \,,
\ee
as the starting point instead of $\eta_u$ and $\eta_d$. The unitary
matrix $U$ that transforms from the basis $\{\eta_{\,-},
\,\eta_{\,+},\,\eta_{\,s}\}$ to that one for the physical meson states 
$\{P_1=\pi^0,\, P_2=\eta,\, P_3=\etp\}$ can then be linearized in the
$\pi^0-\eta$ and $\pi^0-\etp$ mixing angles. An appropriate
parameterization of $U$ reads ($\beta,\psi\propto {\cal O}(\lambda)$,
$\lambda\ll 1$)
\be 
U(\phi,\beta,\psi) \=\left(\begin{array}{ccc}
          1 & \beta +\psi \cos{\phi} & -\psi \sin{\phi} \\
       -\psi -\beta \cos{\phi} & \cos{\phi} & -\sin{\phi}\\ 
       -\beta \sin{\phi} & \sin{\phi} & \phantom{-}\cos{\phi} 
                                \end{array} \right)\,,
\label{matrix}
\ee
with, $U\,U^{\dagger}=1+{\cal O}(\lambda^2)$. It is also of advantage
to introduce isovector and isoscalar axial vector currents   
\be
J^{\mp}_{5\mu}\= \frac1{\sqrt{2}}\, [\ub\gamma_\mu \gamma_5\, u \mp \db\gamma_\mu
    \gamma_5\, d]\,.
\ee
The matrix elements $\langle 0|J^b_{5\mu}|\eta_{b'}\rangle$
($b,b'=-,+,s$) then define a decay matrix 
\be
{\cal F} \=  \left( \begin{array}{ccc}
       f_+ & z f_+ & 0 \\
       z f_+ & f_+ & 0  \\
       0 & 0 & f_s \end{array} \right)\,,
\label{decay}
\ee
which in contrast to \req{currents}, is non-diagonal. The decay
constants $f_+$ and $f_s$  are the basic decay constants in the 
$\eta$-$\etp$ sector while the parameter $z=(f_u-f_d)/(f_u+f_d)$
being obviously of order $\lambda$, occurs in the $\pi^0-\eta (\etp)$
sector. We note in passing that the basic decay constants are
renormalization scale dependent \ci{kaiser}. Ratios like $y$ or mixing
angle are on the other hand scale independent. Since the anomalous
dimension controlling the scale dependence of the decay constants are
of order $\alpha_s^2$ this effect is tiny and discarded here.    

Taking vacuum-particle matrix elements of the current divergences,
one finds with the help of \req{currents}, \req{matrix} and \req{decay}
($a,a'=1,2,3$; $b, b'= -,+,s$) 
\be
\langle 0|\partial^\mu J^b_{5\mu}|P_a\rangle \= {\cal
  M}^2_{aa'}\,U_{a'b'}\, {\cal F}_{b'b}\,,
\label{divergence}
\ee
where ${\cal M}^2=diag[M_{\pi^0}^2, M_\eta^2, M_{\etp}^2]$ is the
particle mass matrix which appears necessarily quadratic here.
Next, I recall the operator relation \cite{witten}
\be
\partial^\mu\, J^a_{5\mu} \= 2 m_a \ab\, i\gamma_5\, a + \omega\,,
\label{anomaly}
\ee
which holds as a consequence of the U${}_{\rm A}$(1) anomaly. The topological
charge density is given by $\omega=\als/4\pi 
G\tilde{G}$ ($G$ denotes the gluon field strength tensor and
$\tilde{G}$ its dual). Inserting \req{anomaly} into \req{divergence}
and neglecting terms of order $\lambda^2$,
one obtains after some algebraic manipulations a set of equations
which can be solved for the mixing parameters
\ba
\sin{\phi} &=& \sqrt{\frac{(M_{\etp}^2-m_{ss}^2)(M_\eta^2-M_{\pi^0}^2)}
                 {(M_{\etp}^2-M_\eta^2)(m_{ss}^2-M_{\pi^0}^2)}}\,,\nn\\[0.2em]
\langle 0|\omega|\eta_+\rangle &=& \frac{f_+}{\sqrt{2}}\, 
                       \frac{(M_{\etp}^2-M_{\pi^0}^2)(M_\eta^2-M_{\pi^0}^2)}
                        {(m_{ss}^2-M_{\pi^0}^2)} \,, \nn\\[0.2em]
y &=& \sqrt{2\, \frac{(M_{\etp}^2-m_{ss}^2)(m_{ss}^2-M_\eta^2)}
                     {(M_{\etp}^2-M_{\pi^0}^2)(M_\eta^2-M_{\pi^0}^2)}}\,,
\label{mix-rel}
\ea 
\be 
   M_{\pi^0}^2 \= m_{++}^2 \= \frac12\, (m_{uu}^2 + m_{dd}^2)\,, \qquad
\label{mass-rel}
\ee
and 
\ba
\beta&=& \frac12\,\frac{m^2_{dd}-m_{uu}^2}{M_{\etp}^2 - M^2_{\pi^0}} +z\,,\nn\\[0.2em]
\psi &=& \frac12\,\cos{\phi}\, \frac{M_{\etp}^2 - M^2_{\eta}}{M_{\etp}^2 - M^2_{\pi^0}}\,
                         \frac{m^2_{dd}-m^2_{uu}}{M_\eta^2 - M^2_{\pi^0}}\,.
\label{iso-rel}
\ea
In addition, the symmetry of the mass matrix forces relations between decay
constants and the anomaly matrix elements
\be
y\=\frac{f_+}{f_s}\=\sqrt{2}\,\frac{\langle 0|\omega|\eta_{\,s}\rangle}
            {\langle 0|\omega|\eta_{\,+}\rangle}\,, \qquad
z\=\frac{f_u-f_d}{f_u+f_d}\= - \frac{\langle 0|\omega|\eta_{\,-}\rangle}
            {\langle 0|\omega|\eta_{\,+}\rangle}\,.
\label{eq:yz}
\ee
Last of all the pion decay constant, $f_\pi$, equals $f_+$ up to
corrections of order $\lambda^2$. The quark mass terms in the above
relations are defined as matrix elements of the pseudoscalar currents 
\be 
m^2_{aa} \= \langle 0|2\, \frac{m_a}{f_a}\,\ab\, i\gamma_5\, a|\eta_{\,a}\rangle\,.
\ee

The quark-flavor mixing scheme can readily be extended to the case of the
$\eta_c$. One then finds for instance~\cite{FKS1} that the charm
admixtures to the $\eta$ and $\etp$ amounts to 0.6 and 1.6
$\%$, respectively. Their corresponding charm decay constants are 
$f^c_{\eta} = -(2.4\pm 0.2)\,\mev$ and $f^c_{\etp} = -(6.3\pm
0.6)\,\mev$. The radiatice decays of the $J/\psi$ into the $\eta$ and
$\etp$ provide a nice test of these results.
\section{$\eta-\etp$ mixing}
The three relations \req{mix-rel}, taken from Ref.~\ci{uppsala}, can 
be used to determine the $\eta-\etp$ mixing parameters for given
masses of the physical mesons and adopting the well-known PCAC result 
\be
m_{ss}^2\= 2 M^2_{K^0} - M^2_{\pi^0}\,,
\label{constraint}
\ee
a result that is also obtained from leading order chiral
perturbation theory. This {\it theoretical} estimate of the $\eta, \etp$
mixing parameters provides 
\be
\phi\= 41.4^\circ\,, \quad f_s\=1.27f_\pi\,,\quad a^2 \equiv \langle
0|\omega|\eta_+\rangle/(\sqrt{2}f_+)\=0.276\,\gev^2\,.
\label{theor-par}
\ee 
It implies a value of $-13.4^\circ$ for the
state mixing angle $\theta=\phi-\theta_{\rm ideal}$ in the
SU(3)${}_{\rm F}$
octet-singlet basis. Note that the theoretical estimate presented
here differs slightly from the one presented in Ref.~\ci{FKS1}. A
{\it phenomenological} determination of the mixing parameters has also been
attempted in Ref.~\ci{FKS1} using experimental data instead of the
theoretical result \req{constraint}. Thus, comparing processes 
involving $\eta$ mesons with those involving $\etp$s one can determine
the mixing angle provided the OZI rule holds. An example is set by the
radiative decays of the $\phi$ meson into the $\eta$ or $\etp$. This decay
proceeds through the emission of the photon from the strange 
quark and a subsequent $s\bar{s}$ transition into the
$\eta_s$ component. Hence, the ratio of $\eta$ and $\etp$ decay widths
reads
\be
\frac{\Gamma[\phi\to \gamma \etp]}{\Gamma[\phi\to \gamma \eta]} \=
\cot^2{\phi} \left(\frac{k_{[\phi\,\gamma\,\etp]}}
                          {k_{[\phi\,\gamma\,\eta]}}\right)^3
\,[1-\kappa_V]\,,
\label{phi-ratio}
\ee
where $k_{[\alpha\,\beta\,\gamma]}$ is the final state's state three momentum
in the rest frame of the decaying meson $\alpha$. The quantity $\kappa_V$ is a
small correction due to vector meson mixing. Other reactions like
the two-photon decays of the $\eta$ or $\etp$ \ci{feld} or the
photon-meson transition form  factors \ci{feld,passek} are sensitive
to the decay constants.    
An interesting piece of information comes from
the radiative decays of the $J/\psi$ into the $\eta$ and $\etp$. According
to Novikov {\it et al}~\ci{nov} the photon is here emitted from the charm quarks
which subsequently annihilate into lighter quark pairs through the
effect of the anomaly. This mechanism leads to the following result
for the ratio of decay widths  
\be
\frac{\Gamma[J/\psi\to \gamma \etp]}{\Gamma[J/\psi\to \gamma \eta]}
 \= \left | \frac{\langle 0|\omega|\etp\rangle}
       {\langle 0|\omega|\eta\rangle}\right |^2\, 
     \left(\frac{k_{[J/\psi\,\gamma\,\etp]}}{k_{[J/\psi\,\gamma\,\eta]}}\right)^3\,.
\label{jpsi-ratio}
\ee   
The analysis of a large class of such reactions leads to the following
set of phenomenological mixing parameters \cite{FKS1}: 
\ba
f_+ &=& (1.07 \pm 0.02)\,f_\pi\,, \quad  f_s = (1.34\pm 0.06)\,f_{\pi}\,, \nn\\
\phi&=&(39.3\pm 1.0)^\circ\,, \qquad a^2\=(0.265\pm 0.01)\,\gev^2\,. 
\label{mix-parameters}
\ea
These values of absorb corrections from higher orders
of flavor symmetry breaking and higher orders of $\lambda$.

Transforming from the quark-flavor basis to the SU(3)${}_F$ basis
by an appropiate orthogonal matrix, one observes the need for two more
angles, $\theta_8$ and $\theta_1$ besides the state mixing angle $\theta$, 
in order to parameterize the constants for the weak decay of a physical
meson through the action of a singlet or octet axial vector current 
\ci{kaiser,FKS1,feld}:
\ba 
f^8_\eta\; &=& f_8 \cos{\theta_8}\,, \qquad f^1_\eta\;\= - f_1
\sin{\theta_1}\,,\nn\\
f^8_{\etp} &=& f_8 \sin{\theta_8}\,, \qquad f^1_{\etp} \= \phantom{-}f_1
\cos{\theta_1}\,.
\ea      
The singlet-octet mixing parameters are related to those for the
quark-flavor basis by
\ba
f_8 &=& \sqrt{1/3 f_+^2 + 2/3 f_s^2}\,, \qquad \theta_8\= \phi -
\arctan{(\sqrt{2} f_s/f_+)}\,, \nn\\ 
f_1 &=& \sqrt{2/3 f_+^2 + 1/3 f_s^2}\,, \qquad \theta_1\= \phi -
\arctan{(\sqrt{2} f_s/f_+)}\,.
\ea
Only in the flavor symmetry limit, i.e.\ if $f_+=f_s$, the three
angles $\theta$, $\theta_8$ and $\theta_1$ fall together. Evaluating
the new parameters, say, from the theoretical set of mixing parameters
\req{theor-par}, one obtains
\ba
f_8&=& 1.18 f_\pi\,, \qquad \theta_8 \= -19.3^\circ\,,\nn\\
f_1&=& 1.09 f_\pi\,, \qquad \theta_1 \= -\;6.9^\circ\,.
\ea
The mixing angles differ from each other and from the
state mixing angle ($-13.4^\circ$) substantially.
\section{Isospin symmetry breaking}
That the character of the approximate flavour symmetry of QCD is
determined by the pattern of the quark masses is a well-known fact 
that has extensively been discussed in the literature for decades. 
Isospin symmetry in particular which would be exact for identical $u$
and $d$ quark masses, holds to a rather high degree 
of accuracy empirically, although the ratio $(m_d-m_u)/(m_d+m_u)$  is
about 1/3, i.e.\ of order 1 and not, as one would expect for a true
symmetry, much smaller than unity \cite{gross}. The violation of
isospin symmetry for pseudoscalar mesons within QCD is usually
estimated as an admixture $\eps_0$ of the flavor-octet $\eta$ state to
the pion. On exploiting the divergences of the axial vector currents,
Gross, Treiman and  Wilczek~\ci{gross} obtained 
\be
\eps_{\,0}\=\frac{\sqrt{3}}{4}\, \frac{m_d-m_u}{m_s - (m_u+m_d)/2}\,,
\label{pcac}
\ee 
a result that also follows from lowest order chiral perturbation
theory~\ci{gasser}. We learn from \req{pcac} that due to the effect of
the U$_{\rm A}$(1) anomaly which is embodied in the divergences of the
axial vector currents, ISB for the pseudoscalar mesons is of the order 
of $(m_d-m_u)/m_s$ instead of the expected order $(m_d-m_u)/(m_d+m_u)$. 
Isospin symmetry is thus partially restored and amounts to only a few
percent. It is therefore to be interpreted  
rather as an accidental symmetry which comes about as a consequence of the
dynamics. For hadrons other than the pseudoscalar mesons the strength
of ISB is not necessarily set by the mass ratio \req{pcac}, for
comments on ISB in the vector meson sector see Ref.~\ci{fritzsch}. 

Defining the isospin-zero admixtures to the $\pi^0$ by
\be
|\pi^0\rangle \= |\eta_{\,-}\rangle + \eps |\eta\rangle + \eps^\prime
|\etp\rangle + {\cal O}(\lambda^2)\,,
\ee
one finds with the help of \req{matrix} and \req{iso-rel}
\ba
\eps &=& \psi + \beta\, \cos{\phi} = \cos{\phi}\, \left[ \frac12\, 
               \frac{m^2_{dd}-m^2_{uu}}{M^2_\eta - M^2_{\pi^0}} +
	       z\right]\,,\nn\\
\eps^\prime &=&\quad\beta\, \sin{\phi}\quad = \sin{\phi}\, \left[\frac12\,
          \frac{m^2_{dd}-m^2_{uu}}{M_{\etp}^2 - M^2_{\pi^0}} +
	       z\right]\,.
\label{angles}
\ea
The $f_u=f_d$ limit of this result, termed $\hat{\eps}=\eps(z=0)$ and
$\hat{\eps}^{\,\prime}=\eps^{\,\prime}(z=0)$ in the following,
coincides with the result reported in \cite{FKS2}. The quark mass term
in \req{angles} may be estimated from the $K^0-K^+$ mass difference 
corrected for mass contributions of electromagnetic origin
\be
m_{dd}^2 - m_{uu}^2 \= 2\,[M^2_{K^0} - M^2_{K^+} - \Delta M^2_{K\,{\rm elm}}]\,.
\ee
According to Dashen's theorem \cite{dashen}, the electromagnetic
correction is given by the corresponding difference of the pion masses 
\be
\Delta M^2_{K\,{\rm elm}}\= M^2_{\pi^0}-M^2_{\pi^+}\,,
\label{pi-elm}
\ee
in the chiral limit. This correction amounts to $-1.26\cdot
10^{-3}\,\gev^2$ and  leads to the estimate $m^2_{dd}-m^2_{uu}=0.0104\gev^2$. 
However, finite quark masses increase $\Delta M^2_{K\,{\rm elm}}$ 
substantially. The exact size of this enhancement is subject to 
controversy. Different authors
\ci{donoghue} obtained rather different values for the electromagnetic
mass splitting of the K mesons. For an estimate of the $z=0$ values of
the $\eta$ and $\etp$ admixtures to the $\pi^0$ I take the average of
the results for  $\Delta M^2_{K\,{\rm elm}}$ quoted in Ref.~\ci{donoghue} and
assign a generous error to the mixing angles in order to take into
account the uncertainites in the electromagnetic contribution to the
Kaon masses. This way I obtain 
\be 
\hat{\eps}\= \eps(z=0) = 0.017 \pm 0.003\,, \qquad
\hat{\eps}^\prime\=\eps^\prime(z=0) = 0.004 \pm 0.001\,.
\label{numerics}
\ee
Due to the different value of the electromagnetic mass correction used now
the value for $\hat{\eps}$ is somewhat larger than the one quoted by
us previously \ci{FKS2,uppsala,feld:99}.

Chao~\ci{chao:89} also investigated $\pi^0 - \eta - \etp$ mixing on
the basis of the axial anomaly but, instead of diagonalizing the mass
matrix, he exploited the PCAC hypothesis. He works in the conventional
singlet-octet basis and assumes that the octet and singlet decay
constants follow the pattern of state mixing, an assumption that has
above been shown to be inadequate and theroretically inconsistent.
Despite this his results on $\eps$ and $\eps^\prime$ agree
with our $z=0$ ones within the errors quoted in \req{numerics}. 

It is elucidating to express the mass terms in \req{angles} by quark masses. 
With the help of the spontanously broken SU(3)$_L\otimes$ SU(3)$_R$
quark model one finds from \req{angles}-\req{pi-elm} 
\be
\hat{\eps} = \sqrt{3}\, \cos{\phi}\, \eps_{\,0}\,,
\ee
with $\eps_{\,0}$ given in \req{pcac}. As we now see there is an
additional factor of $\sqrt{3}\cos{\phi}=1.34$ in comparison with
GTW result \req{pcac}. It would be unity if $\phi=\theta_{\rm ideal}$, 
i.e.\ if the physical $\eta$ and $\etp$ mesons are pure 
flavour octet and singlet states, respectively. The small GTW value of
$\eps_0=0.011$ has its source in the disregard of $\eta-\etp$ mixing
and the use of Dashen's result for the electromagetic Kaon mass
splitting. If the decay constants $f_u$ and $f_d$ differ from each
other the mixing angles may deviate from the values quoted in 
\req{numerics}. This potentially large effect is a source of 
considerable theoretical uncertainty of our understanding of ISB in
the pseudoscalar meson sector. 

It is to be emphasized that ISB as a consequence of $\pi^0-\eta,\,\etp$
mixing, is accompanied by a non-zero vacuum-$\pi^0$ anomaly matrix
element. From \req{matrix}, \req{iso-rel} and \req{constraint}, one
finds 
\be
\langle 0|\omega|\pi^0\rangle \= 2\, \frac{\hat{\eps}}{\cos{\phi}}\,
               \frac{M_{K^0}^2 -M_{\pi^0}^2}{M_{\etp}^2-M_{\pi^0}^2}\,
                    \langle 0|\omega|\eta_{\,+}\rangle\,.
\label{pi-anomaly}
\ee   
The $z$-dependence cancels in this anomaly matrix element. Unavoidably 
the result \req{pi-anomaly} goes along with a strange quark
contamination of the pion.
\section{Phenomenology of ISB} 
In this section I am going to compare experimental results on ISB in
strong interactions with the theoretical expectation.   
 
The decays $\Psi(2S)\to J/\psi P$  are expected to be anomaly dominated
\cite{vol:80} as the radiative $J/\psi$ decays, i.e.\ the $\pi^0/\eta$ 
ratio is controlled by the ratio of the corresponding anomaly matrix
elements:
\be 
\frac{\Gamma[\Psi(2S)\to J/\psi\, \pi^0]}{\Gamma[\Psi(2S)\to J/\psi\,\eta]}
\= \left| \frac{\langle 0|\omega|\pi^0\rangle}{\langle
  0|\omega|\eta\rangle}\right|^2\;
  \left(\frac{k_{[\Psi(2S),J/\psi,\pi^0]}}
    {k_{[\Psi(2S),J/\psi,\eta]}}\right)^3\,.   
\label{ratio-psip}
\ee
The ratio of the anomaly matrix elements
calculated in the described mixing approach (see \req{pi-anomaly}), reads
\be
\frac{\langle 0|\omega|\pi^0\rangle}{\langle
  0|\omega|\eta\rangle} \= \frac{\hat{\eps}}{\cos^2{\phi}}\,, 
\label{ratio-ano}
\ee
where, for simplicity, $M^2_{\pi^0}$ is neglected as compared to
$M^2_{\eta}$ and $M^2_{\etp}$. It is important to realize that there is
no $z$-dependence in \req{ratio-ano}. According to Donoghue and
Wyler~\ci{don-wyl}, possible electromagnetic contributions to the 
transitions  $\Psi(2S)\to J/\psi \,P$ are expected to be strongly 
suppressed. The recent accurate measurement of the branching ratios for 
these decays~\ci{BES}, used in \req{ratio-psip},
\req{ratio-ano}, leads to
\be
\hat{\eps}_{\Psi(2S)} \= 0.031 \pm 0.002\,, 
\label{BES}
\ee
which is rather large as compared with the theoretical estimate 
\req{numerics}. This causes a
problem since the decays $\Psi(2S)\to J/\psi P$  are considered 
as a clean measurement of the $\pi^0-\eta$ mixing angle. 
We don't know what the origin of this discrepancy is. ISB through
mixing seems to be well understood, it is very difficult to obtain a
mixing angle as large as \req{BES} this way. Thus, one may suspect
that either the decays  $\Psi(2S)\to J/\psi \,P$ receive substantial
contributions from other mechanisms than the anomaly (e.g\ from higher
Fock states) and/or there are other sources of ISB within QCD beyond mixing. 

Recently clear signals for ISB and/or charge
symmetry breaking have experimentally been observed in a number of
hadronic reactions \ci{tippens:01,opper:03,steph:03,abdel:03}. However, 
the extraction of the mixing angles from these data is
difficult and model-dependent. 
The experimental ratio of the $\pi^+d\to pp\eta$ and $\pi^-d\to nn \eta$ 
deviates from unity, the charge symmetry result \ci{tippens:01}. On
the basis of a rather simple model that includes $\pi^0 - \eta$ state 
mixing but ignores mixing with the $\etp$, and that takes into account 
a number of corrections such as differences in the meson-nucleon
coupling constants or the proton-neutron mass difference,
the authors of Ref.\ \ci{tippens:01} extracted
a $\pi^0 -\eta$ mixing angle of 
\be
\eps_{\,\pi d} \= 0.026 \pm 0.007\,,
\label{eps(pid)}
\ee      
from their experimental data. State mixing considered in the analysis
performed in Ref.~\ci{tippens:01} 
involves $\eps$ and not just the $z=0$ value, $\hat{\eps}$, that
occurs for instance in the anomaly matrix elements. One may assign the 
difference between the results \req{eps(pid)} and \req{angles} to the
quantity $z$. It then turns out that the small value of 
\be
z_{\,\pi d}\=0.012 \pm 0.010\,,
\ee
i.e.\ a small difference in the individual decay constants
\be
f_d \simeq\, f_+\,(1-z)\,, \qquad  f_u \simeq f_+\, (1+z)\,,
\ee
of about $2\%$, suffices to bring the theory into agreement with the
experimental value \req{eps(pid)}. However, this explanation would
leave the result \req{eps(pid)} unexplained. 

The non-zero forward-backward asymmetry in $np\to d\pi^0$
measured at TRIUMF~\ci{opper:03} is in conflict with charge
symmetry. The phenomenological analysis of this data suffers from 
large ambiguities. A product of the not to well-known $\eta$-nucleon 
coupling constant and the $\pi^0-\eta$ mixing angle controls the
asymmetry. Moreover, there are additional charge symmetry violating 
contributions from the rescattering amplitude of the exchanged pion. 
The latter are practically unknown. Despite this the result presented 
in Ref.~\ci{opper:03} is in agreement with the mixing ideas advocated 
here. For instance, taking the mixing angle \req{numerics} together 
with the small $\eta$-nucleon coupling constant that follows from 
dispersion theory \ci{grein} and is compatible with the generalized 
Goldberger-Treiman relations \ci{feld:99}, one obtains about the same 
value for the product of both as is quoted in Ref.~\ci{opper:03}.  

The cross section for $dd\to {}^4He\,\pi^0$ has not yet been analyzed
theoretically while the COSY measurement of the ratio of the $pd\to
{}^3\,H \pi^+$ and $pd\to {}^3He\, \pi^0$ cross sections provides only
a very weak signal for ISB \ci{abdel:03}.  

Other processes in which ISB occurs are e.g.\ the $\eta (\etp)\to
3\pi$ decays~\ci{leut:96b} or CP violations in $K^0\to \pi\pi$
\ci{ecker}. A somewhat larger value of the $\pi^0 - \eta$ mixing angle
than \req{pcac} seems to required by the data.
With regard of the new experimental result \req{BES}, a revision
of these analyses is perhaps advisable. Recently a new scalar meson 
$D_s^*(2317)$ has been observed \ci{babar} that has an isospin
symmetry violating decay into the $D_s\pi^0$ channel. The branching 
ratio for this decay channel which has not yet been measured, would
again probe the mixing picture of ISB \ci{bardeen}.     
\section{Summary}
A detailed theoretical and phenomenological analysis revealed that the
quark-flavor mixing scheme provides a clean and consistent description
of the mixing of pseudoscalar mesons. On exploiting the divergencies
of the axial vector currents all basis mixing parmeters can be
determined.  It turned out that flavor symmetry breaking manifests
itself differently in the mixing properties of states and decay constants.  
The mixing of the $\eta$ and $\etp$ with the $\pi^0$ induces ISB of
about a few percent. As the comparison with experiment reveals the
exact magnitude of ISB is not well understood as yet.   
Here, more work is clearly needed.

\end{document}